\documentclass[12pt,fleqn]{article} 
\usepackage{graphics} 
\usepackage{cite}
\usepackage{amsfonts} 
\usepackage{a4wide}
\newcommand{\gsim}{\stackrel{\scriptstyle >}{{ }_{\sim}}}
\newcommand{\lsim}{\stackrel{\scriptstyle <}{{ }_{\sim}}}
\newcommand{\mb}{\ensuremath{m_b}}

\newcommand{\Dmb}[1][]{\ensuremath{\Delta\mb^{#1}}}

\newcommand{\Dmf}[1][]{\ensuremath{\Delta m_f^{#1}}}
\newcommand{\tb}[1][]{\ensuremath{\tan^{#1}\!\beta}}

\newcommand{\GeV}{\ensuremath{\,{\rm GeV}}}

\newcommand{\brHtb}{\ensuremath{BR(H^{+} \to t\bar{b})}}
\newcommand{\brHtaunu}{\ensuremath{BR(H^{+}\to \tau^+\nu^-)}}
\newcommand{\brHbb}{\ensuremath{BR(H\to b\bar{b})}}
\newcommand{\brHtt}{\ensuremath{BR(H\to \tau^+\tau^-)}}

\newcommand{\hb}{\ensuremath{h_b}}
\newcommand{\htau}{\ensuremath{h_\tau}}

\newcommand{\Dmtau}{\ensuremath{\Delta m_\tau}}
\newcommand{\mHc}{\ensuremath{m_{H^\pm}}}
\newcommand{\fb}{\ensuremath{\rm \,fb}}
\newcommand{\pb}{\ensuremath{\rm \,pb}}

\newcommand{\CGGJS}{Guasch:1995rn,Coarasa:1998qa,Coarasa:1998ky}
\newcommand{\jaume}{Belyaev:2001qm,Belyaev:2002eq,Belyaev:2002sa}
\newcommand{\logan}{Carena:2001bg,Curiel:2001ad,Herrero:2001yg,Haber:2000kq}
\newcommand{\tauola}{Jadach:1990mz,Jezabek:1992qp,Jadach:1993hs}
\newcommand{\ketevi}{Assamagan:2002ne,Assamagan:2002hz}
\newcommand{\Roy}{Roy:1999xw,Moretti:1999bw,Miller:1999bm}

\begin{document} 
\noindent
\hfill{} 
\begin{tabular}{l} 
MPP-2003-132\\ 
PSI-PR-04-02\\
SHEP-03-36\\
BNL-72033-2004-CP\\
hep-ph/0402212
\end{tabular}
\vspace*{0.1truecm} 
\renewcommand{\thefootnote}{\fnsymbol{footnote}}
\begin{center}
\textbf{\large Determining the ratio of the
$H^{+} \rightarrow \tau \nu$ to $H^{+} \rightarrow t \bar b$ 
decay rates \\[0.25cm]
for large $\tan \beta$ at the Large Hadron Collider}
\vspace*{0.4cm}\\
{\par\centering 
K.A.~Assamagan$^{\,\small{a},*}$, J.~Guasch$^{\,\small{b},*}$, 
S.~Moretti$^{\,\small{c},*}$ and 
S.~Pe{\~n}aranda $^{\small{d},}$}{\footnote{Electronic addresses:
  Ketevi.Adikle.Assamagan@cern.ch, 
jaume.guasch@psi.ch, stefano@hep.phys. ~~~ soton.ac.uk, 
siannah@mppmu.mpg.de.}}
\par 
\vspace*{0.3cm}
{\par\centering 
\textit{$^{\,\small{a}}$Brookhaven National Laboratory, Upton NY 11792, USA}\\
\textit{$^{\,\small{b}}$Paul Scherrer Institut, CH-5232 Villigen PSI,
  Switzerland}\\
\textit{$^{\,\small{c}}$School of Physics and Astronomy,
Southampton University, Highfield SO17 1BJ, UK}\\
\textit{$^{\,\small{d}}$Max-Planck-Institut f\"{u}r Physik,  F\"{o}hringer Ring 6,
    D-80805 M\"{u}nchen, Germany }}
\end{center}
\vspace*{0.3cm}
{\par\centering\textbf{\large Abstract}\\ \par} 
\noindent 
\begin{quotation}\small
We present results on the determination of the observable ratio
$R=\brHtaunu/\brHtb$ of charged Higgs boson decay rates as
a discriminant quantity between Supersymmetric  and
non-Super\-symmetric models. Simulation of measurements of this quantity
    through the analysis of the charged Higgs production process $g b
    \rightarrow t b H^+$ and relative backgrounds in the two above 
    decay channels has been performed in the context of ATLAS. 
    A $\sim 12-14 \%$ accuracy on $R$ can be achieved for $\tan \beta=50\,,
    \mHc=300-500\GeV$ and after an integrated luminosity of $300
    \fb^{-1}$. With this 
    precision measurement, the Large Hadron Collider 
(LHC) can easily discriminate between models for
    the two above scenarios, so long as $\tan\beta > 20$.
  \end{quotation}
  \renewcommand{\thefootnote}{\arabic{footnote}}
\setcounter{footnote}{0}

\vspace*{0.2cm}
\noindent
In this note we investigate the production of charged Higgs bosons in
association with top quarks at the LHC, from the experimental and
theoretical point of view, by studying hadronic ($H^+\to t\bar{b}$) and
leptonic ($H^+\to \tau^+\nu$) decay signatures. The
interest of this investigation is many-fold.
\begin{itemize}
\item The discovery of a charged Higgs boson will point immediately to
  the existence of some extension of the Standard Model (SM).
\item The associated production of a charged Higgs boson with a top quark
  ($pp\to H^+ \bar{t}+X$)~{\cite{Gunion:1994sv,Barger:1994th}} is 
  only relevant at large values of
  $\tb$\footnote{It is in principle
   also relevant at very low values of $\tb$ (say, $\lsim 1$).
   In practise, this $\tb$ regime is excluded in the Minimal Supersymmetric
Standard Model (MSSM) from the
    negative neutral Higgs search at LEP~\cite{:2001xx}.
    Hence, hereafter, we will refrain from investigating the low $\tan\beta$
    case.}, a regime where
  Higgs boson observables receive large Supersymmetric (SUSY) radiative
  corrections. 
\item While SUSY radiative effects might be difficult to discern in the
  production cross-sections separately, they will appear neatly in the
  following relation between the two mentioned channels:
  \begin{equation}
    \label{eq:relation}
    R\equiv\frac{\sigma(pp\to H^+\bar{t} + X \to \tau^+ \nu t + X)}{\sigma(pp\to H^+\bar{t} + X \to t\bar{b} \bar{t} +
    X)}\,\,.
  \end{equation}
\item In fact, in the ratio of~(\ref{eq:relation}), the dependence on the
production mode (and on its large sources of
  uncertainty deriving from parton luminosity, unknown QCD radiative
  corrections, scale choices, etc.) cancels out:
  \begin{equation}
    \label{eq:relation2}
    R=\frac{BR(H^+\to \tau^+\nu)}{BR(H^+\to t\bar{b})}=\frac{\Gamma(H^+\to \tau^+\nu)}{\Gamma(H^+\to t\bar{b})} \,\,.
  \end{equation}
\end{itemize}
From these remarks it is clear that the quantity
$R$ is extremely interesting both experimentally and
theoretically in investigating the nature of Electro-Weak Symmetry Breaking
(EWSB).

In the MSSM, Higgs boson couplings to down-type fermions receive
large quantum corrections, enhanced by \tb. These corrections
have been resummed to all orders in perturbation theory with the help of
the effective Lagrangian formalism  for the $t\bar{b}H^+$
vertex~\cite{Carena:1999py,Guasch:2003cv}. 
The $b$-quark Yukawa coupling, \hb\,, is related to the 
corresponding running mass at tree level by $\hb=\mb/v_1$.
Once radiative corrections are taken into account, due to the breaking of
SUSY, this relation is modified to
$\mb\equiv\hb v_1
\left(1+\Delta\mb\right)$~\cite{Carena:1999py,Guasch:2003cv}, 
where $v_i$ is the Vacuum Expectation Value (VEV)
 of the Higgs doublet  $H_i$ and
$\Dmb$ is a non-decoupling quantity that encodes the leading higher
order effects. Similarly to the $b$-quark case, 
the relation between $m_\tau$ and
the $\tau$-lepton Yukawa coupling, $h_\tau$, is also modified
by quantum corrections, $\Dmtau$. {We} adopt in our
analysis the effective Lagrangian approach by 
 relating the fermion mass to the Yukawa coupling via a generic
$\Dmf$ ($f=b,\tau$),
\begin{equation}
  \label{eq:deffhb}
  h_f=\frac{m_f(Q)}{v_1} \frac{1}{1+\Dmf}=
      \frac{m_f(Q)}{v \cos\beta}\frac{1}{1+\Dmf} \quad
    (v= (v_1^2+v_2^2)^{1/2},  \quad \tan\beta=\frac{v_1}{v_2})\,,
\end{equation}
in which the resummation of all possible $\tb$
enhanced corrections of the type $(\alpha_{s} \tb)^n$ is
included~\cite{Carena:1999py,Guasch:2003cv}. 
The leading part of the (potentially) non-decoupling
contributions proportional to soft-SUSY-breaking trilinear scalar
couplings ($A_f$) can be absorbed in the definition of
the effective Yukawa coupling at low energies and only subleading
effects survive~\cite{Guasch:2003cv}. Therefore,
the expression~(\ref{eq:deffhb}) contains all (potentially) large leading 
radiative effects. The SUSY-QCD contributions to $\Dmb$ are proportional 
to the Higgsino mass parameter $\Dmb\sim\mu$, while the leading SUSY-EW
contributions behave like $\Dmb\sim \mu A_t$~\cite{Guasch:2001wv}. Thus, 
they
can either enhance or screen each other, depending on the sign of $A_t$. 
It is precisely these effects that will allow us to  
distinguish between different Higgs mechanisms of EWSB.
For example, the analysis of 
these corrections in the ratio of neutral Higgs boson decay rates,
$R'={\brHbb}/{\brHtt}$, revealed large deviations from the SM values 
for several MSSM parameter combinations~\cite{Guasch:2001wv}. 
Extensive theoretical analyses of
one-loop corrections to both neutral and charged Higgs boson decays have
been performed in~\cite{\CGGJS,Guasch:1998jc,\jaume,Guasch:2001wv,\logan}. 
We now explore the one-loop MSSM contributions to the ratio of the 
branching ratios ($BR\,$s) of a charged Higgs boson
$H^{\pm}$ in~(\ref{eq:relation2}), which at leading order (and neglecting
kinematical factors) is given by $R=\htau^2/3 \hb^2$ in the large $\tb$
limit. The SUSY
corrections to the $H^+t\bar{b}$ vertex entering the
decay processes $t\to H^+b$ and $H^+\to
t\bar{b}$ have been analysed in~\cite{\CGGJS,Guasch:1998jc},
where it was shown that they change significantly the
Tevatron limits on $\mHc$~\cite{Guasch:1998jc}. They were further explored
in the production process $pp(p\bar{p})\to H^- t\bar{b}$
at LHC and Tevatron in~\cite{\jaume,Guchait:2001pi}, where they were shown to shift
significantly the prospects for discovery of a charged Higgs boson at both 
colliders. 

\medskip
Here, we have performed a detailed phenomenological analysis for the LHC
of charged Higgs boson signatures, by using the subprocess 
$g\bar{b}\to H^+\bar{t}$. The QCD corrections to this
channel are known to next-to-leading
(NLO)~\cite{Plehn:2002vy}. However, we have normalised our
production cross-section to the LO {result}, for consistency with the
tree-level treatment of the backgrounds\footnote{In all the analysis we
  disregard the subleading QCD and SUSY corrections which affect the
  signal and the background, and will take only into account the leading
  SUSY corrections to the signal cross-section, which are absent in the
  background processes.}. In our simulation,
we have let the 
top quarks decay through the SM-like channel
$t\to W^+b$. In the hadronic decay channel of the charged Higgs boson
($H^+\to t\bar{b}$) we require one of the two $W$'s
emerging from the decay chain $H^+\bar t\to (t\bar{b})\bar t
\to (W^+b)\bar b(\bar bW^-)$ to
decay leptonically, to provide an efficient trigger, while the other $W$
is forced to decay hadronically, since this mode provides
the largest rate and in order 
to avoid excessive missing energy. The $\tau$-lepton in the $H^+\to\tau^+\nu$
decay mode is searched for through hadronic one- and multi-prong channels.
In summary, the
experimental signatures of the two production channels under investigation are
($l=e,\mu$):
\begin{eqnarray}
\label{eq:leptonic} 
  pp(g\bar{b})&\to & H^+ \bar{t} \to (\tau^+\nu) \bar{t} \to \tau^+\nu\,
  (jj \bar{b})\,\,,\\
  \label{eq:hadronic}
  pp(g\bar{b})&\to & H^+ \bar{t} \to (t\bar{b}) \bar{t} \to (jj [l\nu]
  b)\, \bar{b} \, (l\nu [jj]\bar{b})\,\,. 
\end{eqnarray}
(In the numerical analysis we always combine the signals 
in~(\ref{eq:leptonic}) and (\ref{eq:hadronic}) with their 
charged-conjugated modes.)

The Monte Carlo (MC) simulation has been performed using  {\tt PYTHIA}
({v6.217})~\cite{Sjostrand:2000wi} for the signal and most of the background
processes. (We have cross-checked the signal cross-section
with~\cite{Plehn:2002vy}.) We have used {\tt HDECAY}\cite{Djouadi:1998yw} 
for the Higgs boson decay rates. One of the background processes (the 
single-top one: see below) has been 
generated with {\tt TopRex} \cite{Slabospitsky:2002ag} with a custom 
interface to {\tt PYTHIA}. We have used {\tt ATLFAST}~\cite{atlfast} for the 
detector simulation. (Further details of the detector can be found
in~\cite{\ketevi}.)  We have adopted the
CTEQ5L~\cite{Lai:1999wy} parton distribution functions in their default
{\tt PYTHIA} {v6.217} setup and we have used
running quark masses derived from the pole values $m_t^{\rm pole}=175\GeV$ and
$m_b^{\rm pole}=4.62\GeV$. The {\tt TAUOLA}~\cite{\tauola} package was 
interfaced 
to the {\tt PYTHIA} event generator for treatment of the  
$\tau$-lepton polarisation. 

\medskip
The leptonic decay channel of the charged Higgs boson provides the
best probe for the detection of such a state at the
LHC. In fact, it turns out that despite the small branching
ratio $BR(H^+\to\tau^+\nu)$, the $\tau$-lepton affords an efficient
trigger to observe this channel. The production rates $\sigma\times
BR(H^+\to \tau^+\nu)\times BR(W\to jj)$ are shown in
Tab.~\ref{tabletau}. The main background processes in this channel
are: top-pair
production with one of the $W$'s decaying into $\tau\nu$ ($gg\to
t\bar{t}\to jj\,b\, \tau\nu\bar{b}$) and $W^\pm t$ associated
production ($g\bar{b}\to W^+\bar{t}\to \tau^+\nu \bar{t}$). 

\begin{figure}[t]
\begin{center}\vspace*{-0.8cm}
\begin{tabular}{cc}
\resizebox{10cm}{!}{\includegraphics{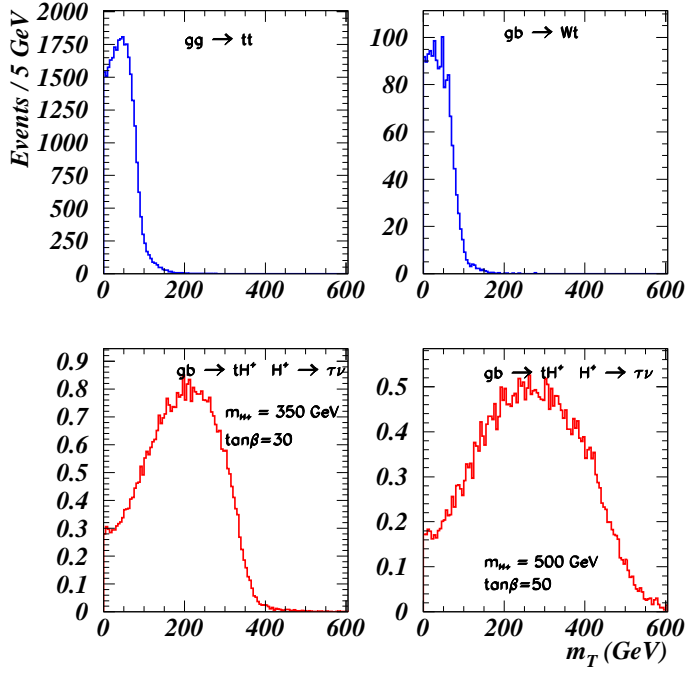}}
\end{tabular}
\end{center}\vspace*{-0.5cm}
\caption{Transverse mass $m_T$ distribution for signal and
  total background taking into account the polarisation of
  the $\tau$-lepton, for an integrated luminosity of $30 \fb^{-1}$. A final 
cut $m_T > 200$~GeV was used for the calculations of the signal-to-background
ratios and for the signal significances.}
\label{fig:transversemass}
\end{figure}

We have used 
the following trigger conditions: hadronic $\tau$-jet ($p_T^{\tau} >
30\GeV$); a $b$-tagged jet ($p_T^{b} > 30 \GeV$) and at least two
light jets  ($p_T^{j} > 30 \GeV$). We apply afterwards a $b$-jet veto
to reject the $t\bar{t}$ QCD background. 
As there is no isolated lepton (electron or muon) in the final state, 
the observation of this channel 
requires a multi-jet trigger with a $\tau$-trigger.
After reconstructing the jet-jet invariant mass $m_{jj}$ and 
retaining the candidates consistent with the $W$-boson mass,
 $|m_W-m_{jj}| < 25\GeV$, the jet four-momenta are rescaled and 
the associated top quark is reconstructed by minimising 
$\chi^2 \equiv (m_{jjb}-m_t)^2$. 
Subsequently, a sufficiently high threshold on the $p_T$ of the
$\tau$-jet is required, $p_T^{\tau} > 100 \GeV$. 
The background events satisfying this cut need 
a large boost from the $W$-boson. This results in a 
small azimuthal opening angle $\Delta\phi$ between the $\tau$-jet 
and the missing transverse momentum, ${p\!\!\!/}_T$. 
For background suppression we then have the cut 
$\Delta\phi({p\!\!\!/}_T,p_T^{\tau}) > 1$. Besides, the missing transverse
momentum is harder for the signal than for the background
while the differences between their
 distributions in azimuthal angle and missing transverse
momentum increase with increasing $m_{H^\pm}$. These 
effects are well cumulated in the transverse mass, 
$m_T = \sqrt{2p_T^\tau{p\!\!\!/}_T\left[1-\cos(\Delta\phi)\right]}$,
which provides good discrimination between the signal and the 
backgrounds, as shown in Fig.~\ref{fig:transversemass}. 
(Further details of this kind of studies are
available in~\cite{\ketevi}.) The discussed set of cuts reduces the
total
 background by six orders of magnitude while the signal is only suppressed
by  two orders.
The production rates and total detection efficiency (including detector
acceptance, $b$- and $\tau$-identification, {pileup} and 
the effect of cuts) are also shown in Tab.~\ref{tabletau} for an integrated
luminosity of $300 \fb^{-1}$. We can see that the signal rates are large
enough to indeed consider  
$H^{+} \rightarrow \tau \nu$ a {\it golden channel} for the $H^{+}$
discovery at large $\tan \beta$.
\begin{table}
\begin{center}
\begin{tabular}{||l|c|c||c|c||}
\hline
& $\mHc=350$ & $\mHc=500$ & $t\bar{t}$ & $W^\pm t$
\\\hline
$\sigma\times BR$ & $99.9\fb$ & $30.7\fb$ & $79.1\pb$ & $16.3\pb$
\\\hline
Events  & 29958 & 9219  & $2.3 \times 10^{6}$ & $4.89 \times
10^{5}$ \\\hline
Events after cuts & 174 & 96 & 17 & 3 \\\hline
Efficiency & 0.6\%& 1\% & $8\times 10^{-6}$ & $6\times10^{-6}$ \\ \hline
$S/B$ & 7.9 & 4.4  \\\cline{1-3}
$S/\sqrt{B}$ & 37.1 & 20.5  \\ \cline{1-3}
Poisson & 23.1 & 14.6 \\\cline{1-3}
\end{tabular}
\end{center}
\caption{The signal and background cross-sections, the number of events before cuts,
  the number of events after all cuts, the total efficiency, 
  the signal-to-background ratios ($S/B$), and the signal significances (Gaussian
  and Poisson) for the detection of the charged Higgs in the $\tau\nu$ channel at
  the LHC, for $300\fb^{-1}$ integrated luminosity and $\tb=50$.}
\label{tabletau}
\end{table}

\medskip
The production rates $\sigma\times
BR(H^+\to t \bar b)\times BR(W^+W^-\to jj l\nu)$ are shown in
Tab.~\ref{tabletb}. The decay mode
$H^\pm \to t b$ has large QCD backgrounds at hadron colliders that come
from $t\bar{t}q$ production with 
$t\bar{t}\rightarrow  Wb\,Wb\rightarrow l\nu b \,jj b$. 
However, the possibility of 
efficient $b$-tagging has considerably improved the situation
\cite{\Roy}. 
We search for an isolated lepton ($p_T^e > 20 \GeV$,  $p_T^{\mu} > 8 \GeV$),
three $b$-tagged jets ($p_T^b > 30 \GeV$) and 
at least two non $b$-jets ($p_T^j > 30 \GeV$). 
We retain the jet-jet combinations whose 
invariant masses are consistent with the $W$-boson mass, 
$|m_W-m_{jj}| < 25\GeV$, then 
we use the $W$-boson mass constraint to find the 
longitudinal component of the neutrino momentum in $W^\pm\to l\nu$, 
by assuming that the missing transverse 
momentum belongs only to the neutrino. Subsequently, 
the two top quarks entering the  $H^+\bar t\to (t\bar{b})\bar t\to (jj [l\nu]
  b)\, \bar{b} \, (l\nu [jj]\bar{b})$ 
decay chain are reconstructed, retaining the 
pairing whose invariant masses $m_{l\nu b}$ and 
$m_{jjb}$ minimise 
$\chi^2 \equiv (m_t-m_{l\nu b})^2 + (m_t - m_{jjb})^2$. 
The remaining $b$-jet can be paired with either top quark to give 
two charged Higgs candidates, one of which leads to a combinatorial 
background. The expected rates for
signal and
background (after the mentioned decays) are shown in Tab.~\ref{tabletb}. 
 (This analysis is presented extensively in~\cite{\ketevi}.)

\begin{figure}[t]
\begin{center}\vspace*{-0.8cm}
\begin{tabular}{cc}
\resizebox{10cm}{!}{\includegraphics{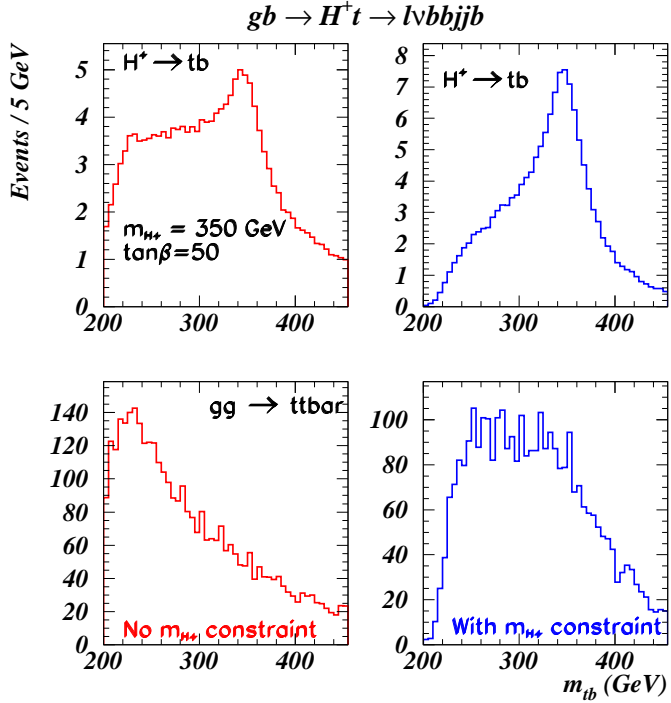}}\\
\end{tabular}
\end{center}\vspace*{-0.6cm}
\caption{The signal and the background distributions for the
  reconstructed invariant mass $m_{tb}$ of $\mHc=350 \GeV$, 
$\tan \beta =50$ and an integrated luminosity of $30 \fb ^{-1}$. Assuming that
the charged Higgs is discovered in the $H^\pm \to \tau\nu$ channel, one can use
$\mHc$ as a constraint to reduce the combinatorial background: this is 
shown on the right plots.}
\label{fig:Hmassrecons}
\end{figure}

At this point, we have two charged Higgs candidates: $t_1 b_3$ or $t_2 b_3$. 
Assuming that the charged Higgs is discovered through the $H^\pm \to \tau\nu$
channel and its mass determined from the $\tau\nu$ transverse mass 
distribution~\cite{\ketevi}, the correct charged Higgs candidate in the $tb$ 
channel can be selected by using the measured $\mHc$ as a constraint. This is 
done by selecting the candidate whose invariant mass is closest to the 
measured charged Higgs mass: $\chi^2 =(m_{tb}-\mHc)^2$.
The signal distribution 
for the reconstructed invariant mass $m_{tb}$ for a charged Higgs boson
weighing $350 \GeV$, with $\tan \beta=50$ and after integrated luminosity 
of $30 \fb ^{-1}$, is shown in 
Fig.~\ref{fig:Hmassrecons}. We can see that for the  
$H^{+}\rightarrow t\bar{b}$ decay some irreducible combinatorial 
noise still appears even 
when the $\mHc$ constraint is included. In addition,
for the background, we have found
that the $\mHc$ constraint reshapes the distributions in
$g g \rightarrow t \bar t X$ in such a way that no improvement in the 
signal-to-background ratio and signal significance is 
further observed. Finally, recall that
the knowledge of the shape and the normalisation of the reshaped background 
would be necessary for the signal extraction. For these reasons, we did not use 
the $\mHc$ constraint for the results shown in this {work}. 
The subtraction of the 
background can then
be done by fitting the side bands and extrapolating in the 
signal region which will be known from the $\mHc$ determination in the 
$H^\pm\to\tau\nu$ channel: however, 
this would be possible only for Higgs masses above 
$300\GeV$ -- see 
Fig.~\ref{fig:Hmassrecons}.  The signal and background results are 
summarised in Tab.~\ref{tabletb} at an
integrated luminosity of $300 \fb ^{-1}$ for different values of
$\mHc$ and $\tan \beta=50$. It is shown that it is
difficult to observe $H^{\pm}$ signals in this channel above $\sim 400 \GeV$, 
even 
with the $\mHc$ constraint. For masses above $\mHc\sim 400\GeV$ the signal 
significance
can be enhanced by using the kinematics of the three-body production 
  process $gg\to H^+\bar{t}b$ \cite{\jaume,\Roy}. 

We assume a theoretical uncertainty of $5\%$ on the branching ratios, $BR\,$s. Previous 
ATLAS studies have shown the residual $gg \to t\bar{t}$ shape and normalisation can be 
determined to $5\%$~\cite{\ketevi}. The scale uncertainties on jet and lepton energies are 
expected to be of the order $1 \%$ and $0.1 \%$ respectively~\cite{\ketevi}. As explained 
above, for $m_{H^\pm} > 300$~GeV, the side band procedure can be used the subtract the 
residual background under the $H^{+}\rightarrow t b$ signal: we
assume also a $5\%$ uncertainty in the background subtraction method. Thus, the
statistical uncertainties can be estimated as $\sqrt{1/S}$. 
The uncertainty in the ratio $R$ are dominated by the
reduced 
knowledge of the background shape and rate in the 
$H^{+} \rightarrow t b$ channel.  The cumulative
results for the two channels
are summarised in Tab.~\ref{tablefinal} at an integrated
luminosity of $300 \fb^{-1}$.  Here,
the final result for the ratio $R$ is
  obtained by correcting the visible production rates after cuts for the
  total detection efficiency in Tabs.~\ref{tabletau} and~\ref{tabletb}
  and by the decay $BR\,$s of the $W$-bosons.
The simulation shows that the above mentioned 
ratio can be measured with
an accuracy of $\sim 12-14 \%$ for $\tan \beta=50$, for
$\mHc=300-500\GeV$ and at an integrated luminosity of $300 \fb^{-1}$. 

\medskip
We turn now to the impact of the SUSY radiative corrections. Their role
is twofold. Firstly, 
by changing the
value of the Yukawa coupling they change the value of the observable
$R$. Secondly, they can change
the value of the production cross-section $\sigma(pp\to H^+\bar{t}+X)$,
hence shifting significantly (by as much as $100 \GeV$) the range of charged
Higgs masses accessible at the LHC~\cite{\jaume}. 

\begin{table}\begin{center}\vspace*{-0.5cm}
\begin{tabular}{||l|c|c||c||}
\hline
& $\mHc=350$ & $\mHc=500$ & $t\bar{t}q$ \\\hline
$\sigma\times BR$ & $248.4\fb$ &  $88\fb$ & $85\pb$
\\\hline
Events  & 74510 &  26389 & $2.55\times 10^{7}$ \\\hline
Events after cuts & 2100 & 784 &  59688
 \\\hline
Efficiency & 2.8\% & 3\% & 0.2\%
 \\ \hline
$S/B$ & 0.035 & 0.013   \\\cline{1-3}
 $S/\sqrt{B}$ & 8.6 & 3.2  \\ \cline{1-3}
\end{tabular}
\end{center}\vspace*{-0.3cm}
\caption{The signal and background cross-sections, the number of events before cuts,
  the number of events after all cuts, total efficiency, $S/B$, and the signal 
  significances for the detection of the charged Higgs in the $tb$ channel at
  the LHC, for $300\fb^{-1}$ integrated luminosity and $\tb=50$.
\label{tabletb}}
\end{table}

\begin{table}\begin{center}
\begin{tabular}{|l|c|c|}
\hline
& $\mHc=350$ & $\mHc=500$ \\\hline
Signals $\tau\nu/tb$ & 174 / 2100 = 0.08 &
96 / 784 = 0.12 \\ \hline
Signals (corrected) $\tau\nu/tb$ &  0.18 & 0.16 \\\hline
Systematics unc. & $\sim 9\%$ & $\sim 9\%$ \\ \hline
Total unc. & 12\% & 14\% \\\hline
Theory & 0.18 & 0.16 \\\hline
\end{tabular}
\end{center}\vspace*{-0.3cm}
\caption{Experimental determination of the ratio~(\ref{eq:relation})
   for $300\fb^{-1}$ and $\tb=50$. Shown are: the signal after cuts, the
   signal after correcting for efficiencies and branching ratios, the
   systematic uncertainty, the total combined uncertainty, and the
   theoretical prediction (without SUSY corrections).\label{tablefinal}}
\end{table}

\begin{figure}[t]
\begin{tabular}{cc}
\resizebox{7cm}{!}{\includegraphics{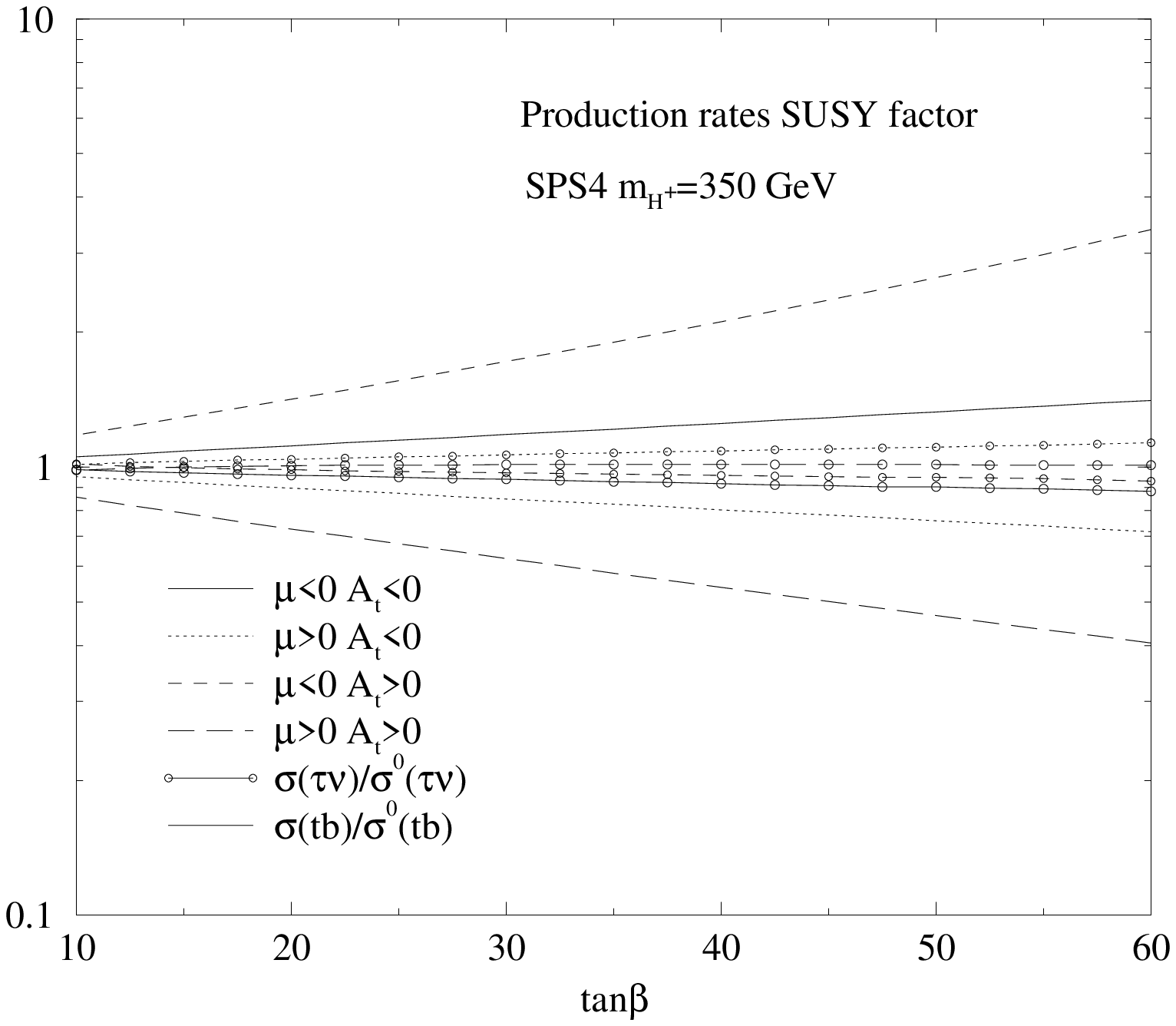}} &
\resizebox{7cm}{!}{\includegraphics{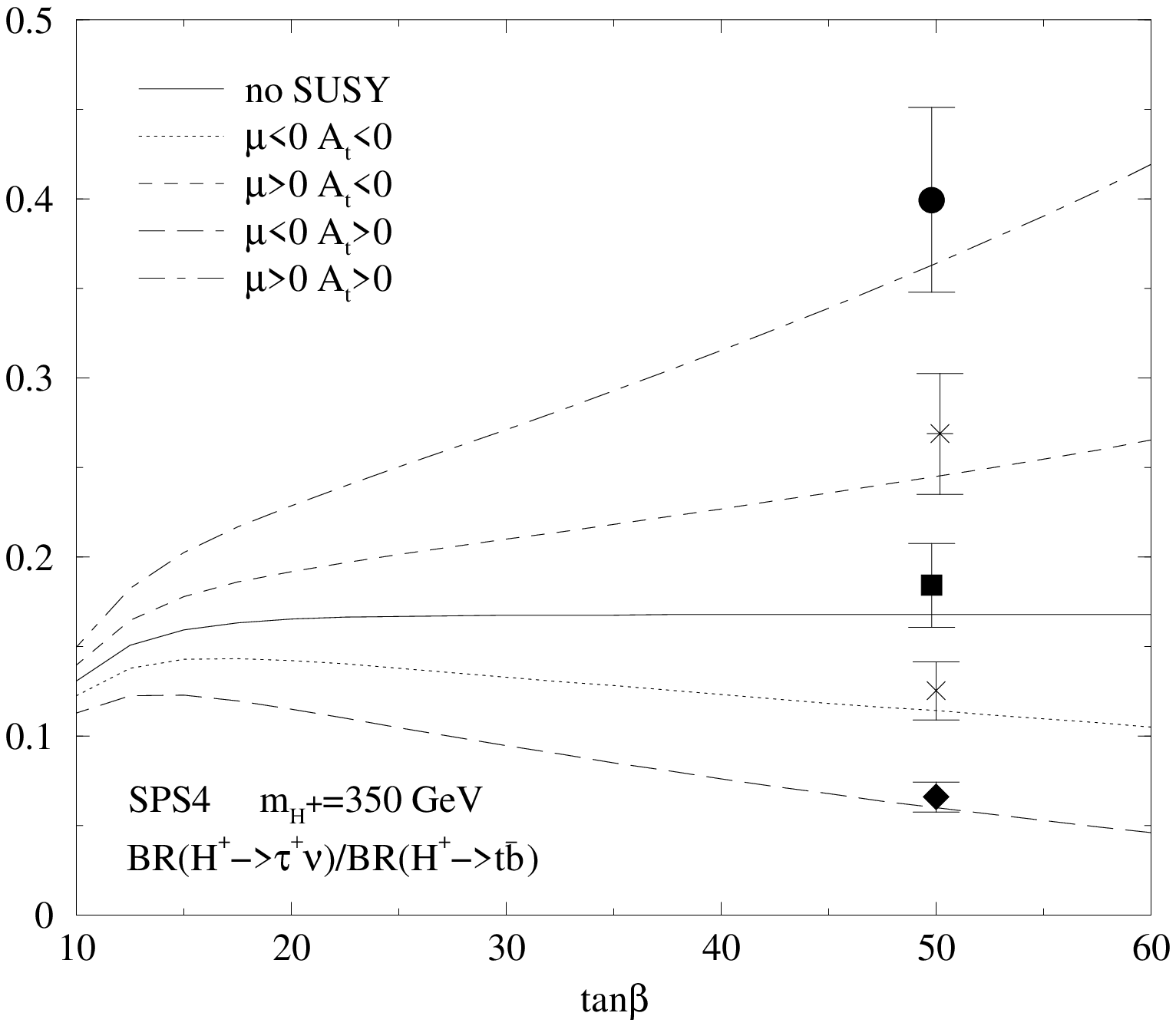}}\\
(a)&(b) 
\end{tabular}
\caption{a) Production rates enhancement/suppression factors
  for the $\tau$ and the $tb$ channels; b) the SUSY correction
  to the rate~(\ref{eq:relation2}). Plots as  functions of $\tb$ for
  $\mHc=350\GeV$ and a SUSY spectrum as in SPS4, but for different
  choices of the signs of $\mu$ and $A_t$. Shown is also the
  experimental determination for each scenario.\label{fig:theory}}
\end{figure}

To explore the second consequence, we rely on the fact that the
bulk of the SUSY corrections to the production cross-section is given by
the Yukawa coupling
redefinition in 
(\ref{eq:deffhb})~\cite{\jaume,Plehn:2002vy}. By neglecting the
kinematic effects, taking the large $\tb$ limit and assuming that the
dominant decay channel of the charged Higgs boson is $H^+\to t\bar{b}$
(large mass limit), we can estimate the corrected production rates. For
simplicity, we show only the contributions to $\Dmb$, since they are the
dominant ones:
\begin{eqnarray}
\sigma^{\rm{corr}}(g\bar{b}\to H^+\bar{t}\to t\bar{b}\bar{t}) &=&
\sigma^{\rm{corr}}(g\bar{b}\to H^+\bar{t}) \times BR^{\rm{corr}}(H^+\to
t\bar{b})\simeq
\frac{\sigma^{0}(g\bar{b}\to H^+\bar{t})}{(1+\Dmb)^2}\,\,,\nonumber\\
\sigma^{\rm{corr}}(g\bar{b}\to H^+\bar{t}\to \tau^+\nu \bar{t}) &=&
\sigma^{\rm{corr}}(g\bar{b}\to H^+\bar{t}) \times
\frac{\Gamma(H^+\to\tau^+\nu)}{\Gamma^{\rm{corr}}(H^+\to t\bar{b})}\nonumber\\
&\simeq&
\frac{\sigma^{0}(g\bar{b}\to H^+\bar{t})}{(1+\Dmb)^2}\times\frac{\Gamma(H^+\to\tau^+\nu)}{\Gamma^{0}(H^+\to t\bar{b})\times\frac{1}{(1+\Dmb)^2}} \,\,.
  \label{eq:correctedrates}
\end{eqnarray}
This very simple exercise shows that the production rate in the $\tau$-channel
 is fairly independent of the SUSY radiative corrections and
therefore the tree-level analysis performed above can (to a very good
approximation) be used for our original purposes. Actually, once we take 
into account kinematical
effects, the $\tau$-channel will receive small (negative)
corrections in the low
charged Higgs mass range. However, in this range, $BR(H^+\to \tau^+ \nu)$
is quite large and one should not fear to loose the signal. Quite the
opposite, the hadronic $t\bar{b}$ production channel receives large radiative
corrections. These corrections can be either positive (enhancing the
signal, and therefore the significance in Tab.~\ref{tabletb}) or negative
(reducing it, possibly below observable
levels)\footnote{Alternative analyses may permit the signal to
  be seen even in this unfavourable case \cite{\jaume}.}. In
Fig.~\ref{fig:theory}a, we show the discussed 
enhancement/suppression factors as a
function of $\tb$ for $\mHc=350\GeV$ and a SUSY mass spectrum defined as
SPS4 of the  {\sl Snowmass Points and Slopes}
in \cite{Allanach:2002nj}, but choosing different scenarios for the
sign of $\mu$ and $A_t$\footnote{The SPS4 spectrum is affected by
moderate SUSY radiative corrections.}.
(It is worth noting that the production rate for the
$tb$-channel can be enhanced by a factor larger than 3 in some SUSY
scenarios, which would enhance significantly the corresponding 
signal thus overcoming the low signal-to-background ratio 
of this channel.)

We now turn our attention to the observable under analysis.
Fig.~\ref{fig:theory}b shows the prediction for the ratio
$R$ as a function of $\tb$ for the
SPS4 scenario with a charged
Higgs mass of $\mHc=350\GeV$. The value of $R$ only depends  on 
$\mHc$ through kinematical factors and the dependence is
weak for $\mHc\gsim 300\GeV$. In this figure, we also show the experimental
determination carried out as before and repeated for each SUSY setup.
From Fig.~\ref{fig:theory}b it is clear that radiative SUSY
effects are visible at the LHC at a large significance. 
In particular, {the $\mu<0$ scenarios can easily be discriminated, while the
  $\mu>0$ ones will be more difficult to establish, due to the lower
  signal rate of the hadronic channel.}
This feature then also allows for a measurement of the sign of the $\mu$
parameter. In contrast, since radiative corrections are independent of the
overall SUSY scale, the observable $R$ cannot provide us with an estimation
of the typical mass of SUSY particles. Nonetheless, the information obtained 
in other production channels (e.g., neutral Higgs bosons or SUSY particles 
direct production) can be used to perform precision tests of the MSSM.

\medskip
To summarise, we have used the observable $R\equiv
\frac{\sigma(pp\to H^+\bar{t} + X \to \tau^+ \nu t + X)}
     {\sigma(pp\to H^+\bar{t} + X \to t\bar{b} t +
    X)}=\frac{\brHtaunu}{\brHtb}\,$ 
 to discriminate between SUSY and non-SUSY Higgs models. This quantity
is a theoretically \textit{clean} observable. The experimental
uncertainties that appear in this ratio have been analysed in details
through detailed phenomenological simulations. In the MSSM, $R$ is affected 
by quantum contributions that do not decouple 
even in the heavy SUSY mass limit. We have quantitatively
shown that an LHC measurement of $R$ can give 
clear evidence for or against the SUSY nature of charged Higgs bosons.

\vspace*{.5cm}

\noindent\textbf{Acknowledgements}\\
KAA,  JG and SM
 wish to thank the Les Houches workshop organisation and
participants for the warm atmosphere of the workshop. JG acknowledges
financial support from the Les Houches workshop organisation and SM from
The Royal Society (London, UK).

\providecommand{\href}[2]{#2}


\end{document}